\documentclass[twocolumn]{aastex631}

\usepackage{comment}
\usepackage{graphicx}
\usepackage{gensymb}

\received{September 21, 2022}
\revised{November 4, 2022}
\accepted{November 6, 2022}

\submitjournal{AJ}

\begin{document}

\title{The on-orbit performance of the Colorado Ultraviolet Transit Experiment ($CUTE$) Mission}

\author[0000-0002-4701-8916]{Arika Egan}
\affiliation{Laboratory for Atmospheric and Space Physics, University of Colorado Boulder, Boulder, CO 80303}
\author[0000-0001-7131-7978]{Nicholas Nell}
\affiliation{Laboratory for Atmospheric and Space Physics, University of Colorado Boulder, Boulder, CO 80303}
\author{Ambily Suresh}
\affiliation{Laboratory for Atmospheric and Space Physics, University of Colorado Boulder, Boulder, CO 80303}
\author[0000-0002-1002-3674]{Kevin France}
\affiliation{Laboratory for Atmospheric and Space Physics, University of Colorado Boulder, Boulder, CO 80303}
\author[0000-0002-2129-0292]{Brian Fleming}
\affiliation{Laboratory for Atmospheric and Space Physics, University of Colorado Boulder, Boulder, CO 80303}
\author[0000-0002-4166-4263]{Aickara Gopinathan Sreejith}
\affiliation{Laboratory for Atmospheric and Space Physics, University of Colorado Boulder, Boulder, CO 80303}
\affiliation{Space Research Institute, Austrian Academy of Sciences, Schmiedlstrasse 6, 8042 Graz, Austria}
\author{Julian Lambert}
\author{Nicholas DeCicco}
\affiliation{Laboratory for Atmospheric and Space Physics, University of Colorado Boulder, Boulder, CO 80303}

\begin{abstract}
We present the on-orbit performance of the Colorado Ultraviolet Transit Experiment ($CUTE$). $CUTE$ is a 6U CubeSat that launched on September 27th, 2021 and is obtaining near-ultraviolet (NUV, 2480 \AA\ -- 3306 \AA) transit spectroscopy of short-period exoplanets.  The instrument comprises a 20 cm $\times$ 8 cm rectangular Cassegrain telescope, an NUV spectrograph with a holographically ruled aberration-correcting diffraction grating, and an NUV-optimized CCD detector. The telescope feeds the spectrograph through an 18\arcmin\ $\times$ 60\arcsec\ slit. The detector is a passively cooled, back-illuminated NUV-enhanced CCD. The spacecraft bus is a Blue Canyon Technologies XB1, which has demonstrated $\leq$ 6\arcsec\ jitter in 56\% of $CUTE$ science exposures. Following spacecraft commissioning, an on-orbit calibration program was executed to characterize the $CUTE$ instrument's on-orbit performance. The results of this calibration indicate that the effective area of $CUTE$ is $\approx$ 19.0 -- 27.5 cm$^{2}$ and that the average intrinsic resolution element is 2.9 \AA\ across the bandpass. This paper describes the measurement of the science instrument performance parameters as well as the thermal and pointing characteristics of the observatory.   

\end{abstract}


\section{Introduction}\label{sec:intro}
The Colorado Ultraviolet Transit Experiment ($CUTE$) is a small satellite currently obtaining near-ultraviolet transit spectroscopy of short-period exoplanets around bright stars. $CUTE$'s spacecraft body, measuring 11.2 cm $\times$ 23.7 cm $\times$ 36.2 cm, is a 6U CubeSat, where a CubeSat is a class of small satellites with external dimensions set by some multiple of a single standardized unit (U) measuring 10 cm $\times$ 10 cm $\times$ 10 cm. $CUTE$ is NASA’s first ultraviolet astronomy CubeSat and the first grant-funded small satellite dedicated to the characterization of exoplanetary atmospheres. The mission was developed at the Laboratory for Atmospheric and Space Physics (LASP) at the University of Colorado Boulder. This paper describes the in-flight instrument performance while a companion paper in this issue, France et al., details the $CUTE$ science and mission overview. Two forthcoming papers will discuss the spacecraft and instrument commissioning (A. Suresh et al. 2023, in preparation) and the data reduction pipeline \cite{Sreejith_2022}. This paper is organized as follows: Section \ref{sec:sci} provides the science motivation for the $CUTE$ mission; Section \ref{sec:missionoverview} provides a small overview of spacecraft testing and the first year of operations; Section \ref{sec:instdesc} describes the spacecraft and payload; Section \ref{sec:instperf} details the instrument performance, including the spectral bandpass and resolution, effective area, background characteristics, and pointing stability; and Section \ref{sec:misops} describes how instrument performance details drive the mission's observing strategies.

\subsection{Science Motivation}\label{sec:sci}
Near-ultraviolet (NUV) transit spectroscopy is a powerful tool for characterizing the upper atmospheric layers of highly-irradiated exoplanets. Planets with short orbital periods of just a few days are bathed in high-energy photons and stellar winds from their host stars, swelling their atmospheres to several planetary radii and potentially past the planet's gravitational boundary (\cite{VidalMadjar2003, Lammer2003, Yelle2004, GarciaMunoz2007, MurrayClay2009, Ehrenreich2011}). Signatures of atmospheric inflation and escape are evidenced in spectroscopic transit light curves created using several strong atomic and ionic absorption features at NUV wavelengths, the depth of which corresponds to the ion's altitude and relative abundance within the atmosphere. For example, Cosmic Origins Spectrograph (COS) NUV observations of the hot Jupiter WASP-12b revealed a pseudo-continuum of metal features throughout the bandpass, with deeper transit depth in \ion{Mg}{2} (2796/2803 \AA) and at several \ion{Fe}{2} wavelengths (\cite{Fossati2010, Haswell2012, Nichols2015}). WASP-121b displayed extended absorption at \ion{Mg}{2}, \ion{Fe}{1} at 2484 \AA\ and in several \ion{Fe}{2} lines, including 2381 \AA\ and 2600 \AA\ (\cite{Sing2019}). HD 209458b has displayed \ion{Fe}{2} at 2370 \AA\ absorbing beyond the planet's Roche lobe (\cite{Cubillos2020}). The shape of a planet's NUV transmission spectrum have the ability to provide constraints on the presence of high-altitude clouds or hazes (\cite{Lothringer2022UV, Wakeford2020, Cubillos2020}). Models using these light curves seek to identify the main drivers of atmospheric outflows and provide estimates for the atmosphere's mass-loss rates.


\begin{figure}
    \centering
    \includegraphics[width = \linewidth]{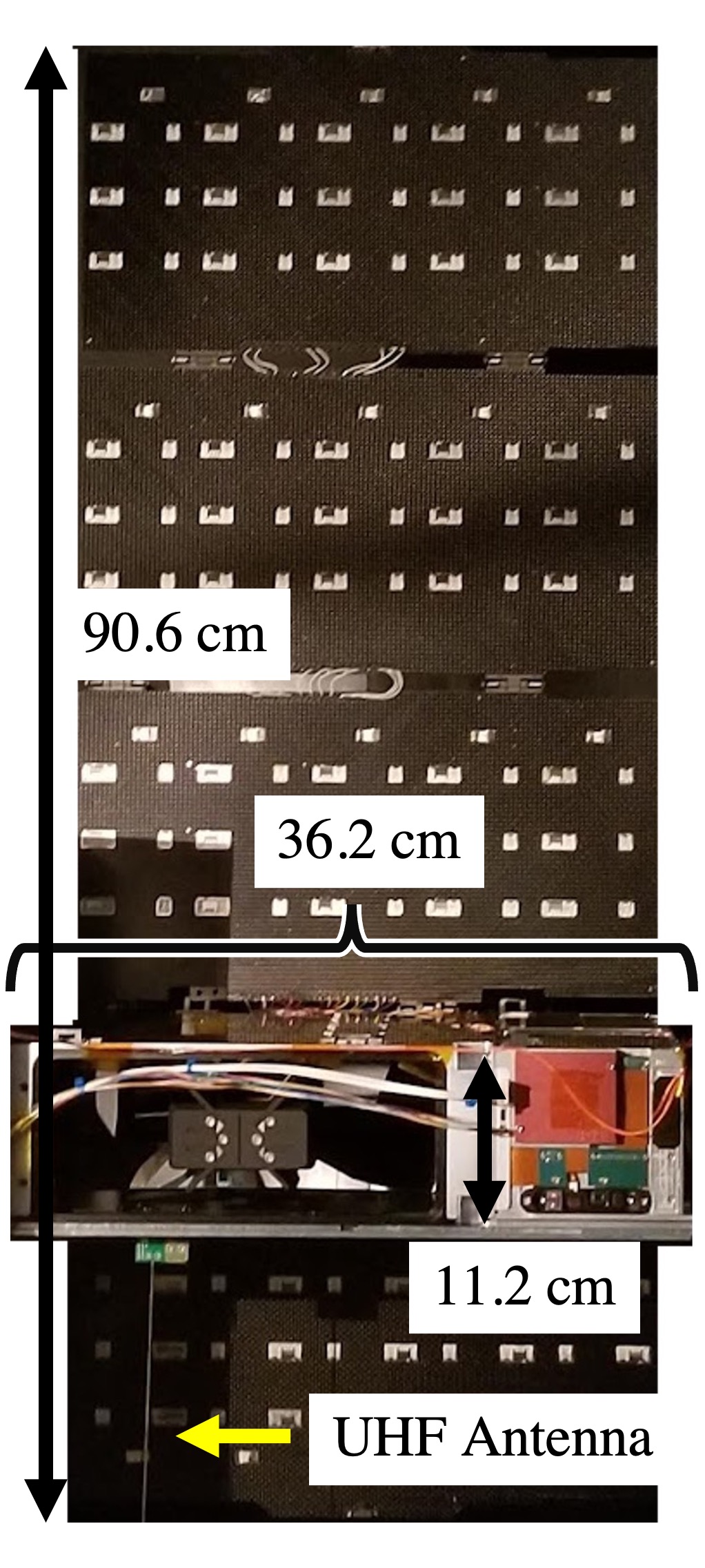}
    \caption{The $CUTE$ spacecraft with solar panels fully deployed and communication cables attached. Solar cells are opposite the telescope boresight. The telescope is visible in the left 2/3 of the spacecraft and the star tracker and avionics are contained in the right 1/3. The startracker has a red ``remove before flight'' panel covering its aperture. The UHF antenna is deployed on the bottom left of the spacecraft chassis and is noted with the yellow arrow.}
    \label{fig:scpic}
\end{figure}

The $CUTE$ mission was designed to observe these NUV absorption features to explore exoplanetary composition and the drivers of atmospheric mass-loss for several known exoplanets: it is observing 6 -- 10 transits of each target to constrain the atmospheric composition properties, identify variability among transits, and provide mass-loss rates for approximately 10 targets over the mission's lifetime.

\subsection{$CUTE$ Mission Overview}\label{sec:missionoverview}
The $CUTE$ CubeSat, shown in Figure \ref{fig:scpic}, was developed, assembled, and tested at LASP from Summer 2017 to Summer 2021. Long lead time items were ordered in the first year, component level calibration and testing occurred in years two and three (including a 10-month delay due to the COVID-19 pandemic), and the majority of assembly and spacecraft testing took place in the year before launch.

The spacecraft was delivered to the Vandenberg Space Force Base on July 21st, 2021 and launched on September 27th, 2021 into an orbit with an average altitude of 560 km, a 97.6$\degree$ inclination, a 10 am local ascending node, and an approximately 95-minute orbital period. Data is downlinked with an S-band radio to the LASP ground station.

Spacecraft and payload commissioning took place shortly after launch through February 2022. We used two main stars to conduct initial payload characterization: $\zeta$ Puppis (HD 66811; O4 star, V = 2.25 mag) and Castor ($\alpha$ Gem, HD 60178; A1 star, V = 1.58 mag). These two stars have International Ultraviolet Explorer ($IUE$) data against which we can calculate $CUTE$'s effective area; they were chosen for their brightness and expected high signal-to-noise ratio that were calculated using $CUTE$ pre-flight effective area estimates and CCD background rates obtained from commissioning exposures. $CUTE$'s on-orbit effective area a representative flux calibrated spectrum are presented in Section \ref{sec:spect}.

$CUTE$ science and dark exposures are typically 300s while the readout time takes an additional 32s; we further command two CCD erasures, or two full transfers using vertical clocking only, before each science, dark, or bias exposure. There are opportunities for up to five 300s exposures per $CUTE$ orbit, though constraints due to solar and lunar keep-out angles, telescope elevation lower limits, and avoidance windows around the north/south poles and South Atlantic Anomaly occasionally reduce that number (France et al. - this issue). Considering the full range of keep-out angles, between 20\% and 30\% of an orbit is used to obtain science and calibration frames. Additional mission operation details will be presented in a forthcoming paper (A. Suresh et al. 2023 - in prep). Full frame images with no processing, called NOPROCs, have 2200 x 515 pixels and are occasionally downlinked to assess the full CCD health. However, due to constrained downlink capacity, the typical science data product is a 2200 x 100 pixel sub-image that is centered on the spectral trace, called a TRIM2D.

$CUTE$ science operations were adjusted post-launch to accommodate damages to the thermoelectric cooler (TEC) and the electronics board that controls both the TEC and the shutter. During thermal vacuum testing, the TEC was damaged. As a secondary payload on a rideshare, it was not possible to replace the TEC and re-test before the delivery date. The CCD now relies on passive cooling (see Figure \ref{fig:ccdtemp}). The damaged TEC likely deposited a small contamination layer on the CCD and nearby optics, potentially degrading the instrument's effective area (Section \ref{sec:spect}).

The TEC and shutter share the same electronics board. While the TEC was damaged pre-launch, the shutter operated nominally and showed no indication of damage. However, during on-orbit payload commissioning, the 12V rail on TEC/shutter electronics board exhibited spikes in its current within a few hours of being powered; these current spikes would trigger the spacecraft's fault protection and reset the spacecraft and payload. The cause of the damage is unclear.

The TEC/shutter electronics board contains a capacitor that will close the shutter whenever the 12V rail loses power, meaning that each spacecraft/payload reset closed the shutter. To prevent the electronics board from continuing to interrupt spacecraft operations, we used a 10 minute pass over the LASP ground station to power the TEC/shutter board, open the shutter, and remove power from the board's 12V line slowly to drain the shutter capacitor to a low enough charge that it would not be able to close the shutter. The shutter is now permanently open. Details about how an open shutter affects $CUTE$ science operations and data reduction are outlined in Sections \ref{sec:back} and \ref{sec:misops}.



\begin{deluxetable}{lc}\label{tab:opt}
\tablecaption{CUTE Optical Summary}
\tablehead{\colhead{Instrument Metric} & \colhead{Value} }
\startdata
Primary Dimensions      & 206 x 84 mm    \\
Primary Radius          & 300 mm         \\
Secondary Dimensions    & 68 $\times$ 26 mm   \\
Secondary Radius        & -129.6 mm       \\
Telescope Focal Ratio   & f/2.6 \\
Telescope PSF FWHM$^{a}$ & 6$\arcsec$ \\
Instrument Focal Ratio  & f/5.5 in cross-dispersion  \\
Slit Dimensions         & 18 \arcmin $\times$ 120$\arcsec$, 60$\arcsec$, or 30\arcsec \\
Grating Dimensions      & 31 $\times$ 31 mm  \\
Grating Radius          & 86.1 mm            \\
Grating groove density  & 1713.4 gr mm$^{-1}$   \\
Mirror Coating          & Al $+$ MgF$_{2}$    \\
Grating Coating         & Bare Al \\
CCD Pixel Size          & 13.5 $\mu$m    \\
CCD Format         & 515 $\times$ 2048 active area  \\
CCD Readout Time        &  32 s         \\
\enddata
\tablenotetext{a}{Point spread function, full width at half maximum. Measured pre-flight only}
\end{deluxetable}

\section{Instrument Description}\label{sec:instdesc}
\begin{figure}
    \centering
    \includegraphics[width = \linewidth, trim={0.25cm 0.25cm 0.25cm 0.25cm},clip]{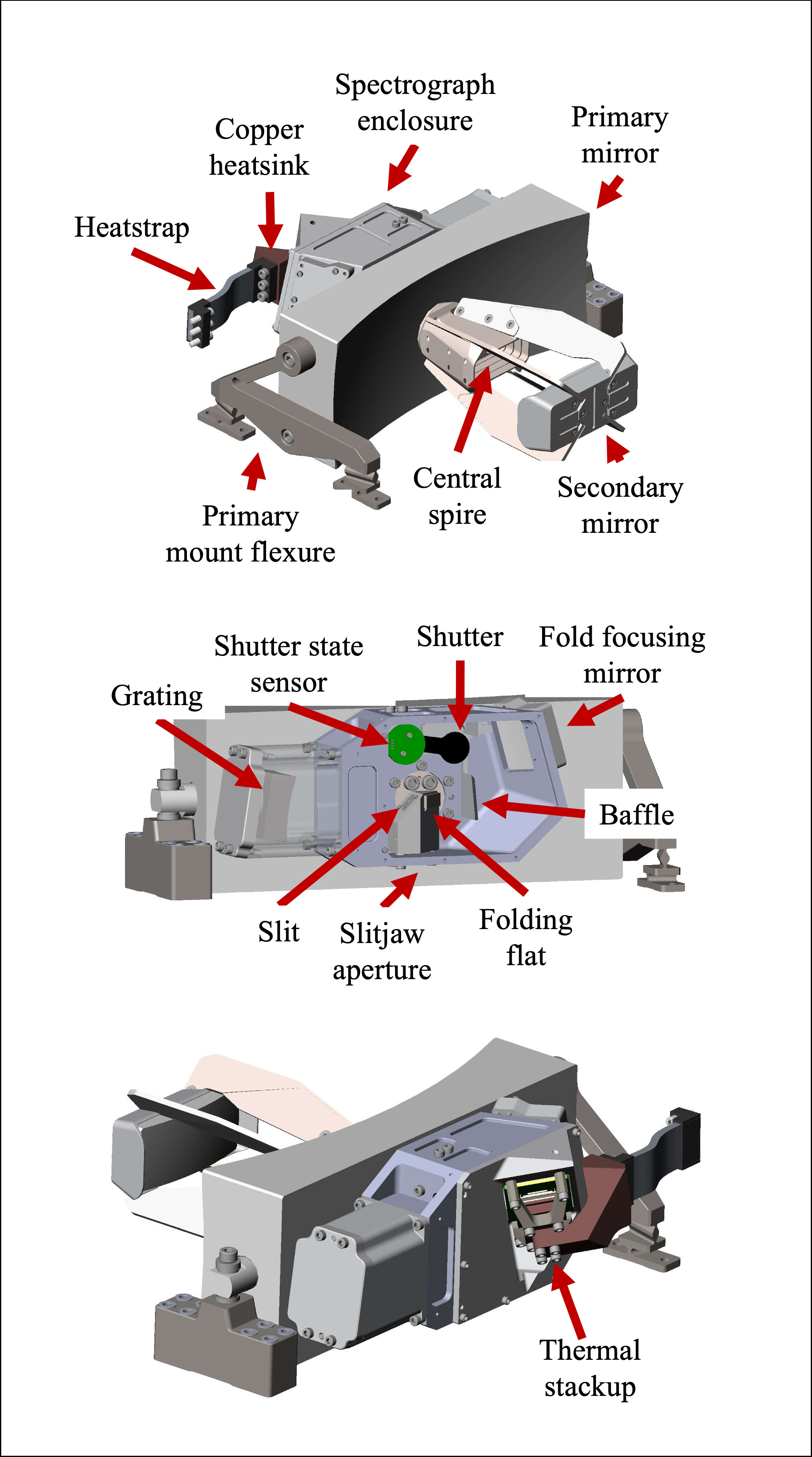}
    \caption{CAD renderings of the $CUTE$ telescope and spectrograph. \textbf{Top}: front view of the Cassegrain telescope. \textbf{Middle}: view of the spectrograph internals. \textbf{Bottom}: view of the fully closed-out spectrograph, including the detector and the thermal strap attached to the spacecraft radiator.}
    \label{fig:instcad}
\end{figure}

The $CUTE$ instrument, shown in Figure \ref{fig:instcad}, is a rectangular Cassegrain telescope provided by Nu-Tek Precision Optical Corporation with an NUV spectrograph and passively cooled CCD. The rectangular primary mirror provides 3$\times$ the surface area of a standard circular mirror fitting into the same volume and was designed to maximize the instrument's light-collecting area in an otherwise small payload volume. The four non-diffractive mirrors are coated in Al $+$ MgF$_{2}$ while the grating is coated in bare Al. The primary mirror serves as the mounting structure for both the secondary mirror and the spectrograph.

\begin{figure*}
    \centering
    \includegraphics[width=\linewidth]{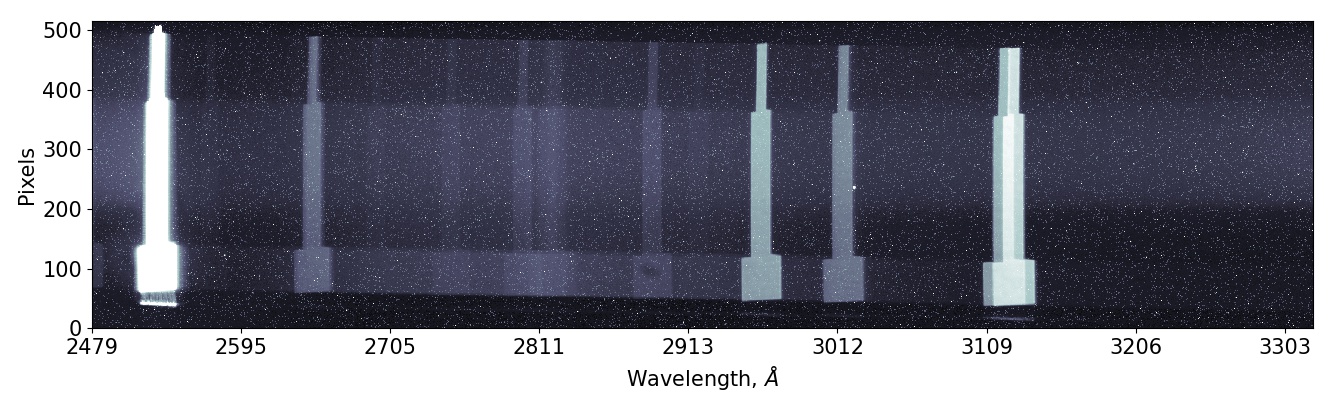}
    \caption{A 2048 $\times$ 515 $CUTE$ CCD image with the slit fully illuminated with a diffuse, uncollimated Hg light source, obtained during thermal vacuum testing. The projected image of the slit is shown at several mercury lines, including 2536 \AA, 2967 \AA, and a doublet at 3125 and 3131 \AA. The exposure time for this image was set to sample the fainter spectral features of the light source; as a result, many pixels in the 2536 \AA\ line reached saturation levels and some side effects of this saturation can be seen in the image.}
    \label{fig:tvacslit}
\end{figure*}

Light from the secondary mirror passes through a ridge-baffled central spire and reflects off of a 45\degree\ fold mirror before reaching a slit at the Cassegrain focus. The slit, manufactured by OSH Stencils, is 18\arcmin\ long in the spatial dimension with three separate sky-projected widths, 30\arcsec, 60\arcsec, and 120\arcsec\, that were chosen to accommodate differently crowded target fields. We have placed the star in the middle of the 60\arcsec section for all observations. A spectral projection of the slit on the CCD is shown in Figure \ref{fig:tvacslit}. After passing through the slit, light is diffracted off of the bare Al coated, holographically ruled, aberration-correcting grating from Horiba-JY and is additionally focused with a cylindrical fold mirror before the spectrum is recorded on a NUV-enhanced back-illuminated e2v CCD42-10 with a 2048 $\times$ 515 active area (CCD details are in Section \ref{sec:ccd} and \cite{Nell2021}). The ridge baffling inside the central spire and additional baffles inside of the spectrograph (Figure \ref{fig:instcad}) mitigate strong scattered light paths, but there are additional signatures of scattered light we identified once in orbit (see Section \ref{sec:back}). Optical design details are provided in Table \ref{tab:opt}.

The CCD is passively cooled; a copper heatsink and thermal strap made of several silver-coated copper wires connect the CCD to a radiator on the side of the spacecraft chassis (\cite{egan2022}). The payload is housed in a Blue Canyon Technologies (BCT) XB1: a 6U spacecraft with 4U housing the payload, 0.5U for instrument electronics, and 1.5U for avionics including the attitude determination and control system (ADCS), batteries, and radios. Four 2U $\times$ 3U solar panels provide power to the spacecraft bus. 


\section{Instrument Performance}\label{sec:instperf}
The $CUTE$ instrument was assembled and tested in LASP vacuum chamber facilities (\cite{France2016}, \cite{egan2020}). Table \ref{tab:perf} details performance parameters between laboratory and in-flight measurements. Bandpass, spectral and spatial resolution, and effective area were all measured with two calibration stars, Castor and $\zeta$ Puppis, and compared against $IUE$ data early in $CUTE$'s commissioning phase. These stars were chosen for their brightness and visibility. Background rates, limiting flux, thermal cycles, and pointing stability were measured in-flight using dark, bias, and science frames from $CUTE$'s first science target, WASP-189b.

\begin{figure*}
    \centering
    \includegraphics[width = \linewidth]{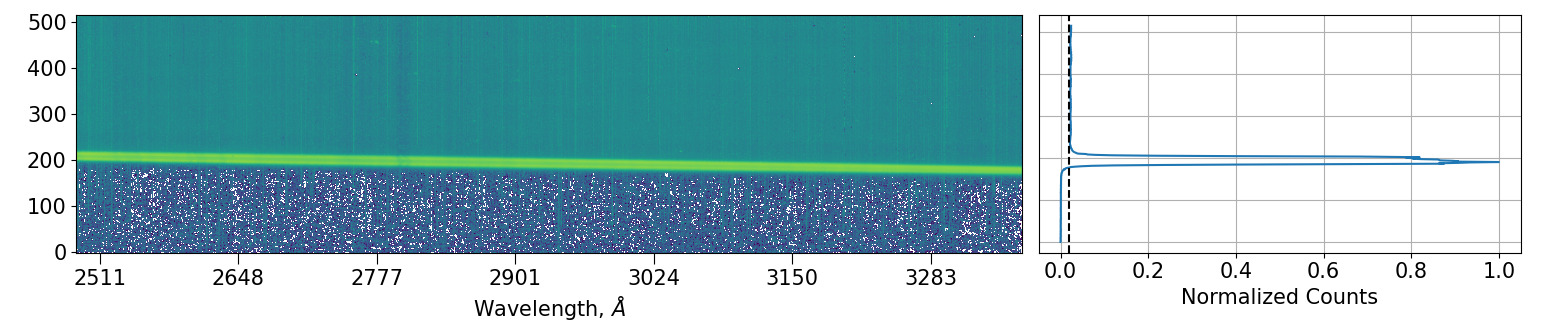}
    \caption{\textbf{Left}: 2048 $\times$ 515 CCD image of Castor's spectra with the the background subtracted. The spectrum is the bright green line in the image's center. Below the spectrum is residual background subtraction noise, and above the spectrum is residual spectral light as the CCD is read out while the shutter remains open. CCD register clocking moves charges to the right and vertical clocking moves the charges down. \textbf{Right}: One-dimensional cross-section of the frame to illustrate the readout streak. The vertical dashed line marks the median value of the readout residual exposure at 2.2\% of the normalized counts cross-section.}
    \label{fig:readout}
\end{figure*}

As $CUTE$ operates without a shutter, CCD pixels remain exposed to light during the 32s readout, and the CCD region above the stellar spectrum contains both background counts and residual spectral counts. An example of this is shown in Figure \ref{fig:readout}. The spectrum is seen in the center of the image, and the residual readout exposure is evident above the spectrum. The normalized one-dimensional cross-section plotted on the right of Figure \ref{fig:readout} illustrates the residual readout exposure. In this example, the increase in counts above the spectrum due to the residual readout is on the order of 2.2\% of the total CCD counts from an observation. For a typical $CUTE$ observation with a 300s exposure time and an average of 150 counts per pixel, the readout time introduces an additional 0.031 counts to each pixel. This is about 0.90\% of the read noise as measured from the blank CCD pixels, and thus the error from the exposed readout is well below the error from other background and noise sources.

\subsection{Spectrograph}\label{sec:spect}
\begin{figure}[htb!]
    \centering
    \includegraphics[width = \linewidth]{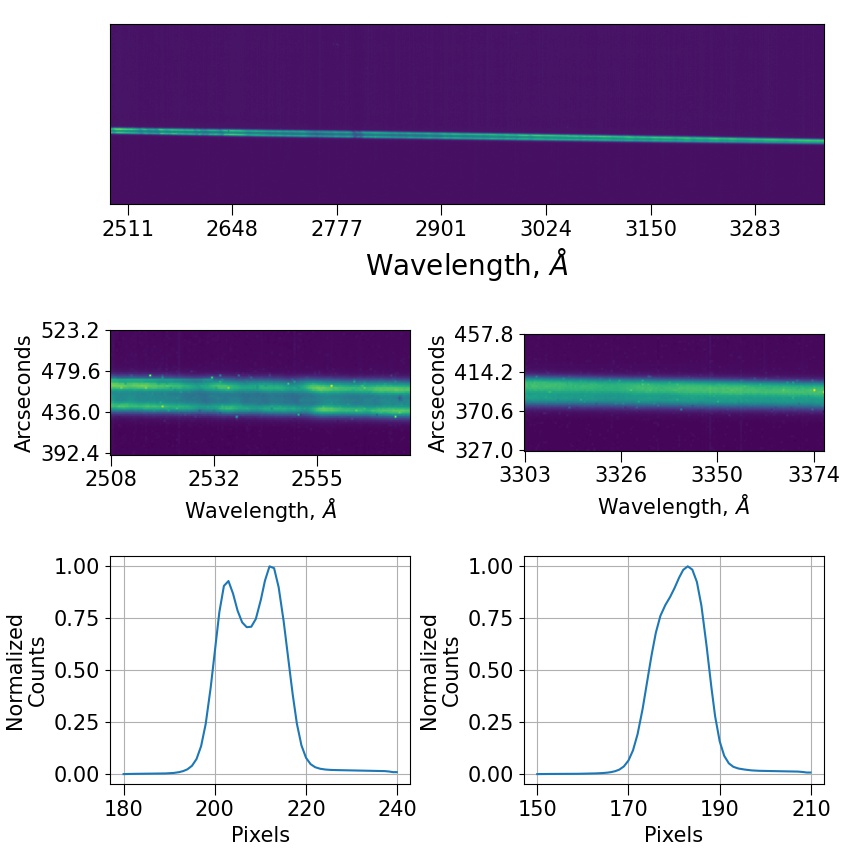}
    \caption{\textbf{Top}: A 2048 x 515 $CUTE$ spectral image of Castor with background effects removed to show the spectrum. \textbf{Middle left}: Zoom of a blue portion of the $CUTE$ spectrum to highlight the out-of-focus lobes. \textbf{Middle Right}: Zoom of a red portion of the $CUTE$ spectrum to highlight the less-out-of-focus lobes. \textbf{Bottom}: One-dimensional cross-sections of the middle panels.}
    \label{fig:2dspeclobe}
\end{figure}

In this section, we characterize $CUTE$'s in-flight spectrograph and present the measured effective area, the bandpass, and the spectral and spatial resolutions. We used $IUE$ observations of Castor to measure the spectral and spatial resolution; $CUTE$'s two-dimensional spectrum of Castor and one-dimensional dispersion profiles are shown in Figure \ref{fig:2dspeclobe}. The tilt of the spectral trace, the asymmetric out-of-focus two-dimensional spectrum, the bandpass, and the dispersion have different values in-flight than measured in the laboratory, indicating shifts in the optical system which likely occurred during launch.

\begin{deluxetable*}{lcc}\label{tab:perf}
    \tablecaption{$CUTE$ Laboratory and measured in-flight performance}
    \tablehead{\colhead{Metric} & \colhead{Laboratory Value} & \colhead{In-flight Value}}
    \tabletypesize{\scriptsize}
    \tablewidth{0pt} 
    \startdata
    Bandpass & 2480 -- 3322 \AA & 2480 -- 3306 \AA \\
    Spectral Tilt & 1.05\degree & 0.85\degree \\
    Intrinsic Spectral Resolution$^{a}$  & 2.1 \AA & 2.9 \AA  \\
    Cross-Dispersion Resolution$^{a}$  & 12.5\arcsec & 30\arcsec \\
    A$_{eff}$ at 2500 \AA & 28.8 cm$^{2}$ & 27.5 cm$^{2}$ \\
    Background Limiting Flux$^{b}$ & N/A & 5 $\times$ 10$^{-14}$ erg s$^{-1}$ cm$^{-2}$ \AA$^{-1}$ \\
    \enddata
    \tablenotetext{a}{Average resolution over the bandpass. In-flight value has ADCS jitter effects removed. }
    \tablenotetext{b}{In 300s, measured on orbit only, evaluated at 3000\AA.}
\end{deluxetable*}

The tilt of the spectral trace is largely a result of the fold focusing mirror's position that was set during the focusing process, though the CCD placement also affects the footprint of the spectrum on the detector. The fold focusing mirror (Figure \ref{fig:instcad}, middle panel) has three adjustment screws on three of the four corners that were used to focus the spectrograph (see \cite{egan2020} for more details). The spectrograph's best focus was found with a spectral trace tilt of 1.08$\degree$; in-flight, the trace's tilt measures 0.85$\degree$.

The spectrograph defocused during launch and now has a double-lobe feature at the blue end that merges into a single lobe at the red end. Despite the change in profile across the detector, the total spectral extraction region to fully capture the extent of the cross-dispersion spectrum remains at a constant value of about 25 pixels across the bandpass for observations with jitter less than 6\arcsec\ RMS, compared to a 13-pixel extraction region measured in the laboratory. For each additional row added to the extraction region, the total noise increases by about 4\%.

\begin{figure}
    \centering
    \includegraphics[width = \linewidth]{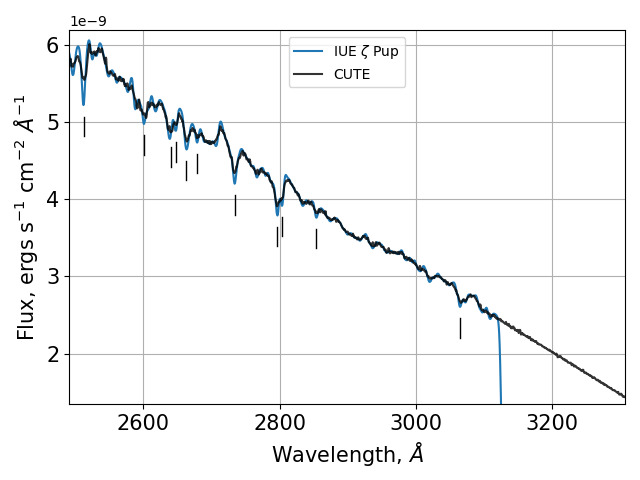}
    \caption{Flux-calibrated spectrum of $\zeta$ Pup. A $CUTE$ spectrum is in black, plotted over the blue $IUE$ spectrum. Black vertical tick marks note the spectral features used to calculate $CUTE$'s wavelength solution. $IUE$ sensitivity falls off beyond 3100 \AA, so a model was fit to the $IUE$ data in order to calculate $CUTE$'s effective area and flux spectrum.}
    \label{fig:fluxcal}
\end{figure}

We used $IUE$ observations of $\zeta$ Puppis to measure the in-flight bandpass; the vertical tick marks in Figure \ref{fig:fluxcal} show the stellar features used to that end. The measured in-flight bandpass is 2480 -- 3306 \AA, a change from the laboratory-measured bandpass of 2479 -- 3322 \AA. The spectrograph's dispersion was also affected by launch; in the laboratory, the bandpass-averaged dispersion was measured between 0.47 \AA/pixel to 0.39 \AA/pixel, and in-flight it is measured to be 0.45 \AA/pixel to 0.35 \AA/pixel. Altogether, the change in trace tilt, the shifted bandpass, the change in dispersion, and the spectrograph's defocus indicate shifts occurring in the optical chain after the Cassegrain focus.

The end-to-end effective area of the optical system was not directly measured prior to launch, but rather calculated using a combination of measured and provided quantities: the telescope's geometric collecting area, optical reflectivities from coating witness samples provided by Nu-Tek for the four non-diffractive reflecting optics, the measured detector quantum efficiency (QE) from \cite{Nell2021}, and laboratory-measured efficiency of the bare Al-coated diffraction grating (\cite{egan2020}).



The in-flight effective area was measured using a combination of $IUE$ flux-calibrated spectra of $\zeta$ Puppis for $\lambda$ $<$ 3100 \AA\ and a linear model for $\lambda$ $>$ 3100 \AA. $IUE$ sensitivity falls off significantly after 3100 \AA\ and therefore our initial calibration program could not directly measure $CUTE$'s effective area beyond 3100 \AA. We instead made the assumption that the slope of the stellar flux from $\lambda \approx$ 2500 -- 3100 \AA\ remained the same for $\lambda \approx$ 3100 -- 3306 \AA\, and used that slope to approximate the stellar flux red-ward of 3100 \AA. We combined the $IUE$ flux and the model to calculate $CUTE$'s effective area. The $IUE$ $\zeta$ Puppis spectra were obtained from the Mikulski Archive for Space Telescopes (MAST) at the Space Telescope Science Institute and can be accessed via\dataset[10.17909/a8wa-vc91]{http://dx.doi.org/10.17909/a8wa-vc91}.

\begin{figure}
    \centering
    \includegraphics[width = \linewidth]{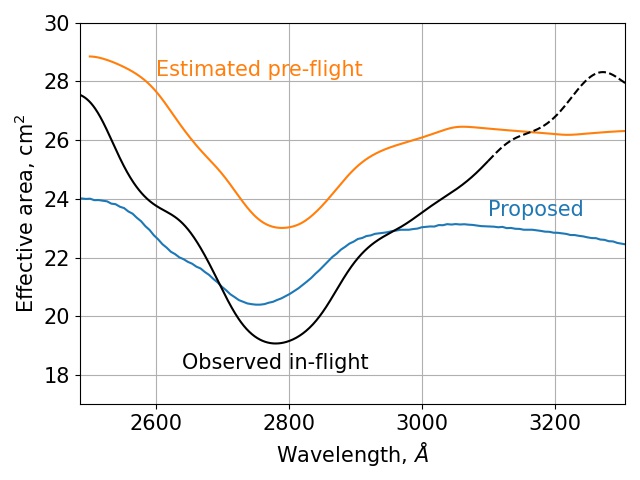}
    \caption{Effective area of the $CUTE$ instrument as proposed (blue), estimated with laboratory measurements (orange), and calculated in-flight and smoothed (black). The in-flight effective area beyond $\lambda$ $\approx$ 3100 \AA\ is estimated based on a model fit to the $IUE$ $\zeta$ Puppis one-dimensional spectrum.}
    \label{fig:effa}
\end{figure}

To arrive at the $CUTE$ effective area in units of cm$^{2}$, the $CUTE$ two-dimensional $\zeta$ Puppis spectrum image was background subtracted, the spectral trace was extracted and summed into a one-dimensional counts spectrum, converted into erg s$^{-1}$ \AA$^{-1}$, and finally divided by the $IUE$ spectra in units of erg s$^{-1}$ cm$^{-2}$ \AA$^{-1}$. The resulting flux-calibrated $CUTE$ spectrum of $\zeta$ Puppis is shown in Figure \ref{fig:fluxcal}. The effective area ranges from 19.0 -- 27.5 cm$^{2}$ across the bandpass and a smoothed effective area curve is shown in Figure \ref{fig:effa}. The in-flight measured effective area is lower than the laboratory-calculated values by a median of 12\%\ across the bandpass. We attribute this change to two possible causes: a contamination layer deposited onto the CCD and optics during the TEC failure (Section \ref{sec:missionoverview}, \cite{egan2022}) and/or atmospheric contamination of the optics during the two month period when $CUTE$ sat in the launch dispenser. However, $CUTE$ science measurements are relative in nature and an individual light curve is made from observations taken within $\approx$24 hours of each other; changes in the instrument's sensitivity between delivery and science operations do not affect $CUTE$ light curve creation, other than the raised noise floor from the TEC failure and launch-induced defocus.



The spectral and spatial resolution were measured using $IUE$ observations of Castor. We define the spatial resolution as the full width at half-maximum (FWHM) of the spectrum in the cross-dispersion direction. This varies by a few pixels across the bandpass (as shown in Figure \ref{fig:2dspeclobe}) with an average of 30\arcsec\ at 3000 \AA\ for observations with jitter less than 6\arcsec\ RMS.

$CUTE$'s spectral resolution as measured from its one-dimensional spectrum comes from the combination of the spectrograph's intrinsic resolution and spectral smearing from spacecraft jitter. Assuming a Gaussian profile for both the spectral line shape and the jitter distribution, the FWHM of each add in quadrature to arrive at the spectral resolution measured directly from $CUTE$ one-dimensional spectra. We convolved $IUE$ spectra of Castor until it matched $CUTE$'s spectra. Using the observation's jitter of 6$\arcsec$ RMS, we arrive at a 2.9 \AA\ intrinsic spectral resolution element.

\subsection{CCD Characterization}
\label{sec:ccd}

The detector has a 2200 $\times$ 515 pixel format with the following layout for each row: 51 true overscan pixels, 50 blank pixels, 2048 active pixels, 50 blank pixels, and 1 null element (a readout artifact used in testing to simplify waveform memory implementation in the FPGA), as can be seen in the full frame images in Figure \ref{fig:scatlightfull}. The 100 blank pixels are not able to sample dark current. We use the horizontal blank and overscan pixels to measure the read noise and bias levels and implement a 5-pixel buffer to avoid charge transfer effects at the active/blank pixel boundary. For example, a full 2200 $\times$ 515 CCD image will have on one side a 50 $\times$ 515 blank region, and a 40 $\times$ 505 sub-region is used to calculate bias and read noise levels. The same 5-pixel buffer enacted on the 102 $\times$ 515 blank $+$ overscan side produces a 92 $\times$ 505 region. The height of these regions is reduced to 90 pixels for a TRIM2D frame. The read noise is considered to be the RMS of these sub-regions and the bias level is calculated from the median. The CCD has a single output channel; register clocking moves charges from the blue end to the red end of the detector, and vertical clocking moves charges from the top to the bottom of the images shown in Figures \ref{fig:tvacslit}, \ref{fig:readout}, and \ref{fig:2dspeclobe}.

Much of the CCD characterization took place in the laboratory and is detailed in \cite{Nell2021}. Quantities including gain, photon-response non-uniformity (PRNU), and non-linearity were not able to be measured in flight due to the time and budget constraints of a suborbital-class mission. In-flight flat fields cannot be obtained as there is no onboard calibration lamp and true flats cannot be obtained with a dispersing element. $CUTE$ science and calibration targets fill the pixel wells to about 5\% of full capacity, levels where non-linearity is not a concern. PRNU, gain, and any dust effects are accounted for by (1) the flux calibration and (2) the nature of light curves being fundamentally a relative measurement.

\subsection{Background}\label{sec:back}
\begin{figure}
    \centering
    \includegraphics[width = \linewidth]{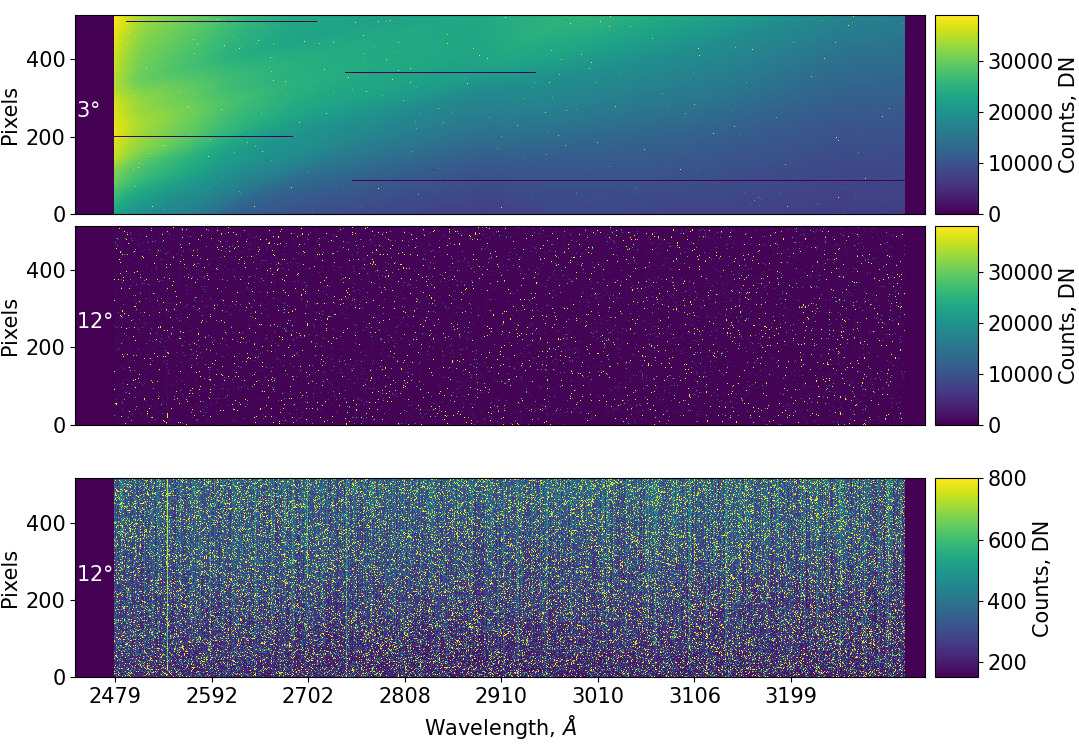}
    \caption{Two 2200 $\times$ 515 $CUTE$ frames of different telescope elevation angles, 3$\degree$ and 12$\degree$, to demonstrate the scattered light feature present at low telescope elevation angles. The elevation of each exposure is marked in the left-side overscan region. The top and middle frames compare the 3$\degree$ and 12$\degree$ elevation angles with the same colorbar. The bottom frame shows the 12$\degree$ frame with a colorbar scaled to the frame. Typical frame features are evident on the 12$\degree$ frame: an increase in counts from the bottom to the top of the frame and hot pixel streaks due to the 32s CCD readout time. The horizontal dark bars in the top image are missing data packets that were unsuccessfully downlinked.}   \label{fig:scatlightfull}
\end{figure}

A single dark frame has contributions from dark noise, read noise, detector bias, and the sky due to the unshuttered spectrograph. The background exhibits additional variation dependent on (1) the CCD's temperature-dependent dark rate and (2) a scattered light feature with a brightness that is correlated with the telescope's elevation angle; an extreme example of this scattered light feature is shown in Figure \ref{fig:scatlightfull}. Both dark and bias frames have some level of contribution to all of the above phenomena.

\begin{figure}
    \centering
    \includegraphics[width = \linewidth]{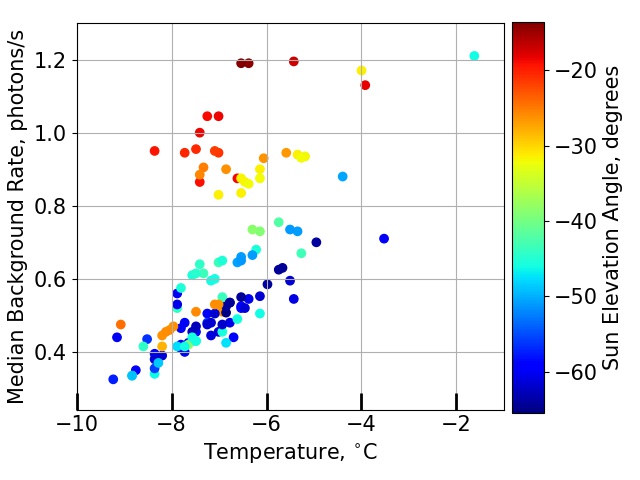}
    \caption{$CUTE$ CCD dark frame background rates for the detector, plotted against detector temperature and colored by the elevation of the Sun with respect to the telescope boresight. In general, higher CCD temperatures produce higher background count rates. Background rates are additionally influenced by the position of astronomical bodies like the Sun as exemplified here, as well as the Moon and Earth. At the time of writing, we are still exploring sources of contribution to background rates.}
    \label{fig:darknoise_slat}
\end{figure}

\begin{figure}
    \centering
    \includegraphics[width = \linewidth]{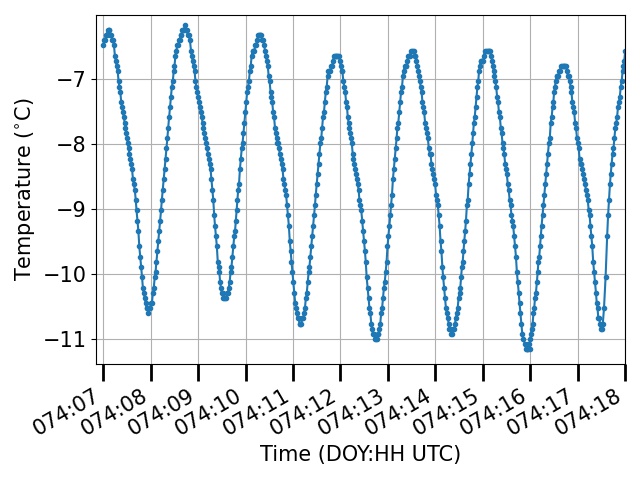}
    \caption{A typical CCD temperature profile over several $CUTE$ orbits.}
    \label{fig:ccdtemp}
\end{figure}

Figure \ref{fig:ccdtemp} shows the detector's periodic temperature cycle over several orbits, cycling between $\approx$ $-$11\degree C and $-6$\degree C. These temperature swings are evident in the average background frame count rates, shown in Figure \ref{fig:darknoise_slat}. The background count rate increases as temperature increases, which is the expected dark rate behavior of the detector. However, the background count rate additionally varies at similar temperatures due to scattered light entering the spectrograph; the angles between the telescope and the Earth's limb, the Sun, and the Moon all have an effect on the level of scatted light present in background frames. An example is shown in Figure \ref{fig:darknoise_slat} where the Sun's elevation angle is correlated with increased background counts. In CCD regions that are more prone to scattered light (e.g. the left side of the CCD as shown in the top panel of Figure \ref{fig:scatlightfull}),  background count rates differ on the order of about 0.15 photons s$^{-1}$ pixel$^{-1}$ when telescope elevation is maintained above 10\degree.

Background subtraction for a single science frame must consider the orbital and thermal environment of the exposure. The temperature- and pointing-dependent nature of the background (e.g. Figure \ref{fig:darknoise_slat}) means that each science frame must have tailored calibration frames. Efforts are underway to model each pixel's value as a function of both its intrinsic behavior and its environment in order to create individual background frames for a given science exposure, based on the frame's exposure conditions; these efforts will be described in detail in Egan et al. 2023 $-$in prep and we currently use a combination of median-combined dark and bias frames with similar telescope pointings and CCD temperatures to approximate the background in each science frame. 

Finally, we present $CUTE$'s background flux limit. After a science frame is background-subtracted, we define a background region below the spectrum that has the same size as the frame's spectral extraction region and calculate the background flux limit, or minimum flux below which a source cannot be detected above the background. The counts in the background region are converted to flux and then averaged for 67 $\times$ 300s WASP-189 science frames. The background flux limit is $\approx$ 5 $\times$ 10$^{-14}$ erg s$^{-1}$ cm$^{-2}$ \AA$^{-1}$. Background subtraction and handling of hot pixels are discussed in more detail in \cite{Sreejith_2022}.

\subsection{Pointing Stability}\label{sec:pointing}

\begin{figure}
    \centering
    \includegraphics[width = \linewidth]{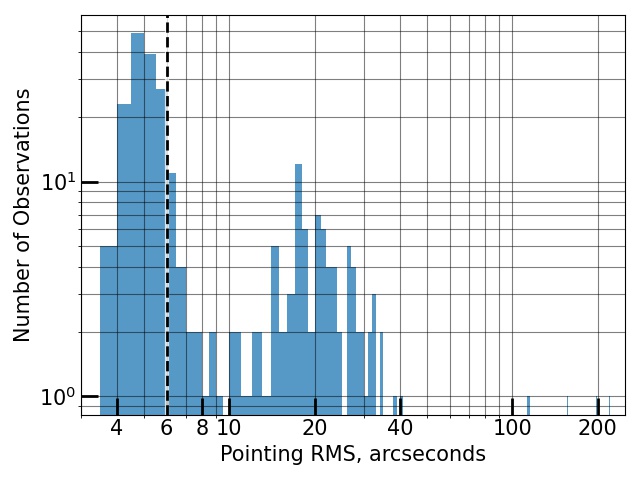}
    \caption{Jitter histogram for 254 $\times$ 300s science exposures from the WASP-189b campaign. The pointing sample rate is 5s for each 300s observation. Histogram bin widths are 0.5\arcsec\ for jitter $<$ 10\arcsec\ and 1\arcsec\ for jitter $>$ 10\arcsec. A vertical dashed line indicates the cutoff used to eliminate low signal-to-noise observations from light curve analysis.}
    \label{fig:pointhist}
\end{figure}

The quality of $CUTE$ spectra is strongly influenced by the spacecraft pointing jitter. Stability about the commanded coordinates can vary between 3\arcsec\ and more than 20\arcsec\ and thereby smear the spectrum across a greater CCD area, increase the spectral extraction region, and reduce the signal-to-noise of the one-dimensional spectrum. Significant jitter can cause vignetting of the telescope point-spread-function by the slit mask and render those data unusable for transit spectroscopy. We currently use a 6\arcsec\ jitter cutoff for science frames to undergo data reduction and observations with higher jitter are not used \cite{Sreejith_2022}. Figure \ref{fig:pointhist} shows a histogram of ADCS jitter in all of the WASP-189 science frames which include complete jitter telemetry; jitter is less than 6\arcsec\ for about 56\% of $CUTE$ observations. In our preliminary data reduction and light curve creation, we remove frames with higher jitter to eliminate low signal-to-noise observations; this cutoff will be honed as we continue to refine our data reduction pipeline (\cite{Sreejith_2022}).




\section{Mission Operations}\label{sec:misops}
$CUTE$'s standard mission operation sequence involves observing a given science target for 6$-$10 transits. Each visit's duration is centered on the planet's mid-transit time and lasts for a span of time equal to five times the transit duration in order to establish an out-of-transit stellar baseline pre- and post-transit. $CUTE$ target transits last between 3 and 6 hours, or about 2 to 4 $CUTE$ orbits, while the duration of each visit is typically less than 24 hours.

As discussed in Section \ref{sec:instperf}, $CUTE$ data quality has pointing and temperature dependencies: (1) CCD exposures are captured without a shutter, meaning that calibration frames are not truly ``dark'' and include sky background; (2) CCD dark rates scale with the thermal changes present due to the passive cooling method; (3) there is a pointing-dependent scattered light feature present in all frames. We create observation plans that take these into account which are detailed below.

The roll angle for a given science target observing campaign is set with two considerations. First, we prioritize keeping the CCD as cool as possible by orienting the radiator panel into space and away from the Earth and Sun. Once that angle is set, we additionally check the target field to make sure no bright stars other than the target are within the slit; this has been the case for all science observations obtained since launch. The roll angle is maintained for all calibration frames taken adjacent to a given set of science exposures. To capture scattered light features from the Sun, Moon, or Earth limb that may appear in science frames (Figure \ref{fig:darknoise_slat}), dark and bias calibration frames are planned to occur at approximately the same orbital position as science frames with a pointing offset from the target star of 0.75$\degree$; this has an additional benefit of obtaining a calibration frame in a similar thermal state.

Fully characterizing the background is a continued effort. Background trends related to scattered light and telescope pointing did not become apparent until a sufficient number of science and dark frames from the WASP-189 observing campaign were analyzed. Modeling efforts are underway to model each pixel's behavior in its environment to better produce calibration frames for background subtraction (Egan et al. 2023 -- in prep.).

\section{Conclusion}\label{sec:summary}
$CUTE$ is currently obtaining NUV transit spectroscopy of short-period exoplanets around bright stars. We have detailed the NUV payload's spectroscopic performance as calculated within the first six months of calibration and science observations, including the spectral and spatial resolution, the effective area, and the limiting flux level. We have demonstrated stable pointing for about 56\% of our observations. Science operations will continue through June 2023, during which we will conduct additional calibration campaigns to measure time-dependent sensitivity, as well as continue to hone our understanding of the background. $CUTE$ is expected to reenter the atmosphere within 3 years from launch.

{\bf Acknowledgments:} $CUTE$ was developed and operated with the support to two NASA/APRA awards to the Laboratory for Atmospheric and Space Physics at the University of Colorado Boulder, NNX17AI84G and 80NSSC21K1667s.  A. G. S. was supported by a Schr\"{o}dinger Fellowship through the Austrian Science Fund (FWF) [J 4596-N] and additionally acknowledges financial support from the Austrian Forschungsf\"orderungsgesellschaft FFG project 859718 and 865968. The $CUTE$ team acknowledges the numerous invaluable discussions with colleagues excited about ultraviolet transit science and the potential to do science with small satellites. The $CUTE$ team wishes to specifically recognize the amateur radio operator community for hosting numerous telemetry tracking tools that have improved the mission's ability to recover from faults and understand long-term spacecraft trends much more efficiently than would have been otherwise possible.

\bibliography{bibtex}{}
\bibliographystyle{aasjournal}

\end{document}